\documentclass[journal=jpclcd,manuscript=letter]{achemso}

\usepackage[version=3]{mhchem} 

\author{Cyril Jean}
\affiliation{
Sorbonne Universit\'es, UPMC Univ Paris 06, UMR 7588, 
Institut des NanoSciences de Paris, F-75005, Paris, France 
}
\alsoaffiliation{
CNRS, UMR 7588, Institut des NanoSciences de Paris, F-75005, Paris, France 
}
\author{Laurent Belliard}
\email{Laurent.Belliard@upmc.fr}
\affiliation{
Sorbonne Universit\'es, UPMC Univ Paris 06, UMR 7588, 
Institut des NanoSciences de Paris, F-75005, Paris, France 
}
\alsoaffiliation{
CNRS, UMR 7588, Institut des NanoSciences de Paris, F-75005, Paris, France 
}

\author{Thomas W. Cornelius}
\author{Olivier Thomas}
\affiliation{%
Aix-Marseille Universit\'e, CNRS UMR 7334, IM2NP, F-13397 Marseille Cedex, 
France
}%

\author{Maria Eugenia Toimil-Molares}
\author{Marco Cassinelli}
\affiliation{%
GSI Helmholtz Centre for Heavy Ion Research, D-64291 Darmstadt, 
Germany
}%
\author{Lo\"ic Becerra}
\affiliation{
Sorbonne Universit\'es, UPMC Univ Paris 06, UMR 7588, 
Institut des NanoSciences de Paris, F-75005, Paris, France 
}
\alsoaffiliation{
CNRS, UMR 7588, Institut des NanoSciences de Paris, F-75005, Paris, France 
}
\author{Bernard Perrin}
\affiliation{
Sorbonne Universit\'es, UPMC Univ Paris 06, UMR 7588, 
Institut des NanoSciences de Paris, F-75005, Paris, France 
}
\alsoaffiliation{
CNRS, UMR 7588, Institut des NanoSciences de Paris, F-75005, Paris, France 
}

\title{Direct Observation of Propagating Gigahertz Coherent Guided 
Acoustic Phonons in Free Standing Single Copper Nanowires}


\begin{document}


\begin{tocentry}

\includegraphics[scale=0.1]{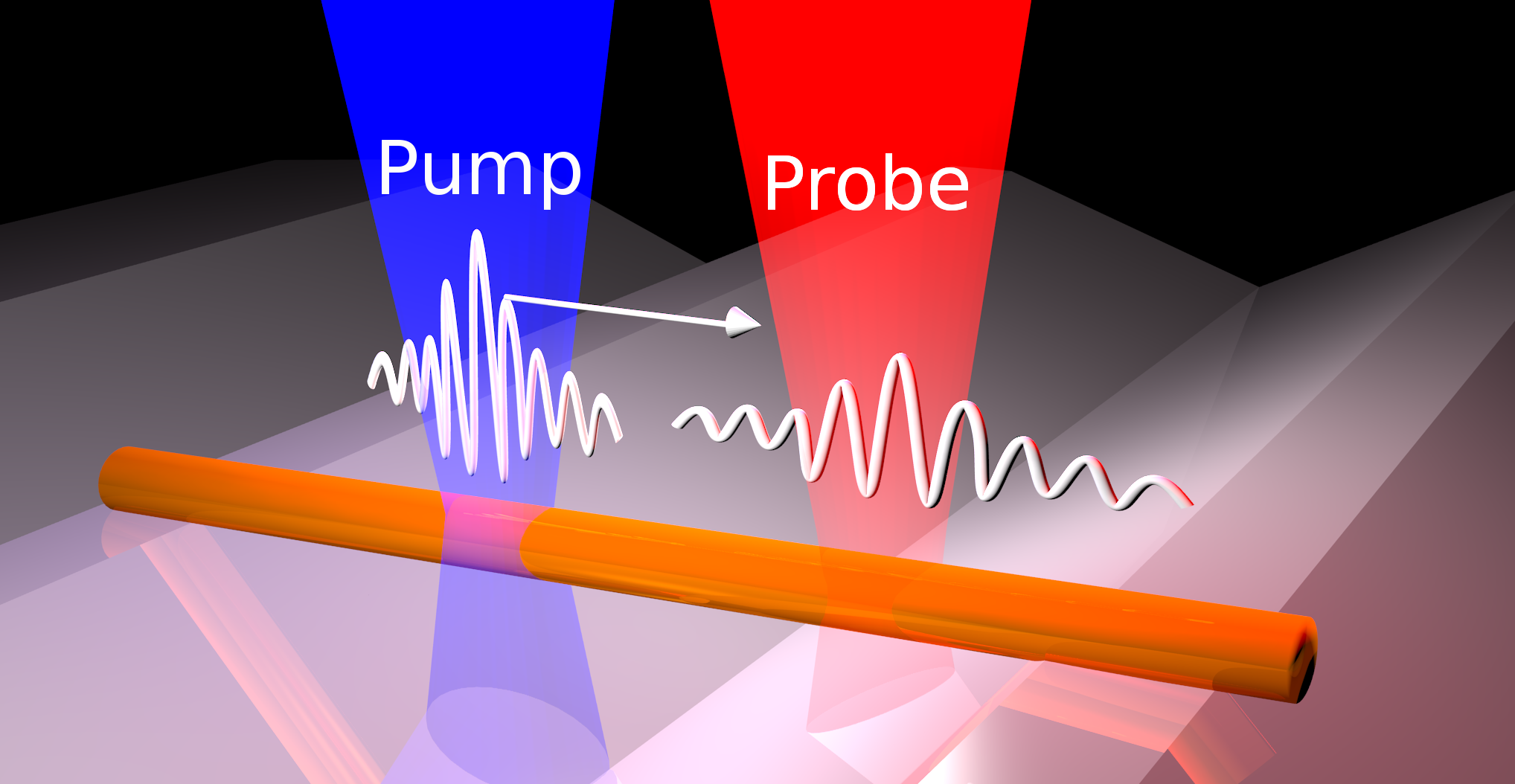}

\end{tocentry}

\begin{abstract}We report on gigahertz acoustic phonon 
waveguiding in free standing single copper nanowires studied by femtosecond 
transient reflectivity measurements. The results are discussed on the basis 
of the semi-analytical resolution of the Pochhammer and Chree equation. 
The spreading of the generated Gaussian wave packet of two different modes 
is derived analytically and compared with the observed oscillations of the sample 
reflectivity. These experiments provide a unique way to independently obtain
geometrical and material characterization. This direct observation of coherent
 guided acoustic phonons in a single nano-objet is also the first step towards 
nanolateral size acoustic transducer and comprehensive studies of the thermal 
properties of nanowires.
\end{abstract}

During the last decade, nanoscale confinement has stimulated wide fundamental 
and technological interests in various fields such as photonics \cite{Maier}, 
electronics \cite{chang2007single}, chemistry \cite{Novo2008} or biology 
\cite{cancer,Hirsch2003}. The impact of size reduction down to the nanoscale
on acoustic response also attracts considerable attention \cite{Major2014}. Fundamental 
motivations include testing the validity of the classical continuum theory of
 elasticity at the
nanoscale \cite{Liang2005,Juve2010}. A better understanding of the phonon 
behavior in nanostructures is also crucial to design MEMS nanoresonators 
\cite{Feng2007}. Confinement deeply modifies the acoustic dispersion relations 
compared with the bulk counterpart. Such modifications strongly influences 
the thermal \cite{Mingo,Boukai2008} and electronic \cite{Bannov} properties of 
the nanostructures. Time resolved optical spectroscopy is now considered a 
powerful tool to address phonon properties in single nanoparticles. In such an
approach, thermal expansion induced by fast lattice heating created by 
femtosecond laser absorption, is optically detected in transmission or 
reflection geometry in far or near field 
\cite{Thomsen,siry2003picosecond,Vertikov1996}. Since the pioneer work of van 
Dijk \textit{et al.} \cite{VanDijk2005}, which investigated the dynamic response 
of single gold nanospheres, a large variety of materials and particle shapes 
have been investigated including nanostructures \cite{Burgin, Bienville2006, 
Amziane2011}, nanowires 
\cite{Staleva,Major2013,staleva2009coupling}, 
nanorings \cite{Kelf}, nanocubes \cite{Staleva2}, nanorods \cite{cardcoreshell,zijlstra} 
or dimer nanoparticules \cite{Jais}. These investigations revealed that the 
system's vibration eigenmodes are strongly correlated with different 
parameters like size, shape, material and coupling with the surroundings. 
Recently, to circumvent the drawback linked to the breathing mode's huge 
damping rate, which occurs due to the interaction between the nanostructure 
and the substrate, free standing nanowires have been investigated 
\cite{Major2013}. The main advantage of this geometry lies in the obtention of
resonators with high quality factors, thus allowing a better understanding 
of their elastic properties \cite{Belliard2013,Ristow2013}. In parallel guided 
acoustic phonons inside nanowires emerge as promising
candidates for nanoacoustic wave generation with nanoscale spotsize, which
could prove useful to design nanodevices for three-dimensional noninvasive 
ultrasonic imaging \cite{Lin2006} with nanometer resolutions. Confined 
\cite{Mounier2014} and propagating \cite{Mounier} acoustic waves have been 
reported recently in a single microfiber (with diameter $>30~\mathrm{\mu m}$). 
However, up to now guided phonons modes propagation in nanowires has
been adressed only for a bundle of $75~\mathrm{nm}$ GaN 
nanowires \cite{Mante2013} and consequently the results were very sensitive to the inhomogeneous 
broadening of the acoustic features due to averaging on nanowires of different 
dimensions. 

In this letter we provide evidence of the propagation of gigahertz coherent 
guided acoustic phonons in single free standing nanowires. 
Beyond the intrinsic novelty of this direct
 experimental observation on a single nano-object, the generation and detection
 of nanoacoustic waves emerge as a useful characterization tool. We first show
 that the experimental investigation of confined modes only allows the
 determination of the characteristic sound velocities $v_L$ and $v_T$\footnote{
$v_L = \sqrt{E(1-\nu)/(\rho(1+ \nu)(1-2\nu))}$ is the longitudinal sound velocity,
 : $v_T = \sqrt{E/(2\rho(1+ \nu))}$ is the transverse sound velocity. $E$, $\nu$ and $\rho$ are
the Young modulus, the Poisson ratio and the density of the material respectively.}, 
the radius $\mathrm{a}$ of the wire has to be determined by an other
 characterization method such as Scanning Electron Microscopy (SEM). We then 
analyze the generation and propagation of coherent guided phonons in single 
copper nanowires. We are able to follow two phonon modes along the nanowire axis.
We first observe the propagation of a gaussian wave packet corresponding to the
propagation of a pure radial breathing mode of frequency around 
$15.6~\mathrm{GHz}$ which exhibits a parabolic dispersion curve. A pulse, 
characteristic of an expansionnal mode associated with a linear dispersion 
curve is also observed. The observed oscillations of the sample reflectivity are 
compared to the predicted behavior in an infinite cylinder.

\begin{figure}
  \includegraphics[scale=1]{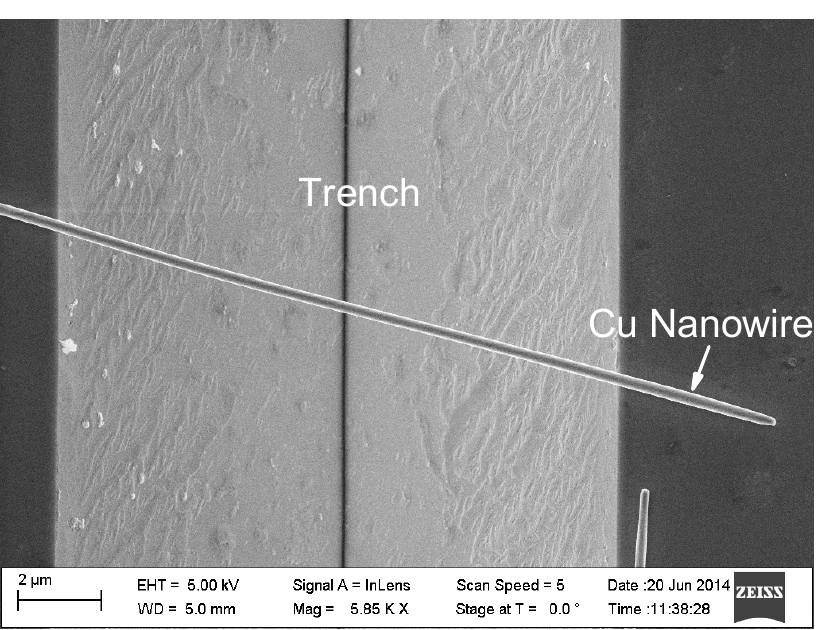}
  \caption{Scanning electron microscope image of a $200~\mathrm{nm}$ 
diameter copper nanowire placed across pyramidal trenches fabricated by 
lithography and anisotropic silicon etching.}
  \label{fig1}
\end{figure}

Since the last decade, many works have been devoted to time resolved 
spectroscopy on supported nanowires. More recently, free standing geometries
have demonstrated a better acoustic confinement.
Our polycrystalline copper nanowires are prepared by electrodeposition in etched ion-track 
membranes as described elsewhere \cite{Belliard2013,molares2001single}. 
In order to reduce the energy dissipation through the silicon substrate, 
the wires are dispersed on a silicon wafer structured with periodic trenches 
(Fig.~\ref{fig1}) fabricated by lithography and anisotropic silicon etching 
\cite{Belliard2013}. Ultrafast pump-probe 
spectroscopy experiments are performed using a mode-locked Ti:sapphire 
(MAI-TAI Spectra) laser source operating at $800~\mathrm{nm}$ with
a pulse duration of $100~\mathrm{fs}$ at a repetition rate of 
$78.8~\mathrm{MHz}$. The pump beam is modulated at $1.8~\mathrm{MHz}$ to 
perform synchronous detection on the sample reflectivity. Both pump and 
probe beams are focused by an objective with a $NA = 0.9$ and are 
normally incident on the sample. The probe beam is fixed on a XY piezoelectric 
stage such as it is laterally positionned with respect to the 
pump beam. To avoid scattered light coming from the pump beam, a two-color 
expriment is performed by doubling the pump frequency 
($\lambda = 400~\mathrm{nm}$) with a nonlinear cystal. A dichroic filter located
in front of the diode system suppresses the light of the pump beam, 
its power is reduced around $300~\mathrm{\mu W}$ and the power of the 
probe beam does not exceed $30~\mathrm{\mu W}$. With such experimental 
conditions, we stay in the thermoelastic regime, the acoustic signal 
and the optical reflectivity remain stable during all the average processing.
The reflectivity from the sample is measured by an avalanche photodiode and 
analyzed with a lock-in amplifier. A maximum pump-probe
time delay equal to $12~\mathrm{ns}$ is achieved using a mobile reflector 
system mounted on a translation stage. 

We investigate nanowires, whose diameter $D$ is around $200~\mathrm{nm}$ and
 whose total length $L$ exceeds $5~\mathrm{\mu m}$, in the frame of the 
elasticity theory of waves developped by Pochhammer \cite{Pochhammer} 
and Chree \cite{Chree}. Our cylinders are considered as infinite given their 
large aspect ratio $L/D > 25$. Due to diffraction limitation, 
the pump laser spot size 
is larger than the nanowires diameter, leading to homogeneous dilatation. 
We thus consider no azimutal dependance of the modes in the following. 
Zero radial stress at the surface is the boundary condition for a free 
standing cylinder. Applying these conditions
 yields the following dispersion equation \cite{royer2000elastic}

\begin{eqnarray}
2P_n\left(Q_n^2+K^2\right)\mathrm{J_1}(P_n)\mathrm{J_1}(Q_n) - 
\left(Q_n^2-K^2\right)^2
\mathrm{J_0}(P_n)\mathrm{J_1}(Q_n) \nonumber\\
- 4K^2P_nQ_n\mathrm{J_1}(P_n)\mathrm{J_0}(Q_n) = 0
\label{eigenvalue}
\end{eqnarray}

\begin{figure}
  \includegraphics{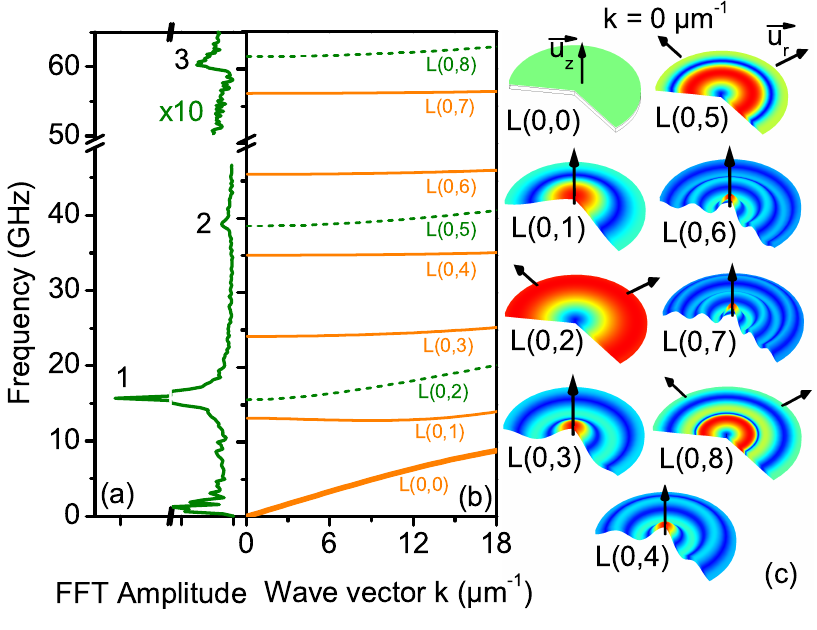}
  \caption{(Color online) (a) : FFT of an experimental oscillatory signal on
an autosuspended $200~\mathrm{nm}$ diameter copper nanowire. One can observe
the first three breathing modes. (b) : numerically calculated dispersion relation
of the nine first longitudinal modes of a $200~\mathrm{nm}$ diameter copper
 nanowire. (c) : Mode shapes of the first eigth non trivial longitudinal 
eigenmodes calculated at
 infinite wavelength ($K = 0 $) using 2D axisymmetric finite elements method (FEM). 
The Young modulus, Poisson ratio and density have been set at 
$110~\mathrm{GPa}$, $0.35$ and $8700~\mathrm{kg \cdot m^{-3}}$ respectively, in 
agreement with usual values for polycrystalline copper\cite{ledbetter1974elastic}. 
The modes $\mathrm{L(0,2)}$, $\mathrm{L(0,5)}$ and 
$\mathrm{L(0,8)}$ exhibit purely radial displacement whereas the 
displacement field of the other six modes is purely axial. Colors 
represent the magnitude of the displacement : a deep blue color corresponds 
to a nodal area of the mode shape when the red color means maximum magnitude. Black arrows indicate the displacement field direction.}
  \label{fig2}
\end{figure}

\noindent where $\mathrm{J_0}$ and $\mathrm{J_1}$ are respectively 
the spherical Bessel function of the first kind of order zero and one. 
The natural number $n$ is the order of the longitudinal axially symmetric
mode, $\mathrm{L(0,n)}$, according to Silk and Bainton naming convention 
\cite{silk1979propagation}. $P_n^2= (X_n/\beta)^2-K^2$ and 
$Q_n^2= X_n^2-K^2$. $X_n$ is the reduced pulsation 
$X_n = \omega_n a / v_T$, $K = ka$ is the reduced wave number and 
$\beta$ is a material parameter which exclusively depends of the Poisson
ratio $\nu$ of the material : $\beta^2 = (v_L/v_T)^2=(2-2\nu)/(1-2\nu)$. 
The dispersion curves of the first nine dilatational modes 
of a $200~\mathrm{nm}$ diameter copper nanowire with a circular 
cross section are calculated numerically and reproduced in 
Fig.\ref{fig2}b. At infinite wavelength ($K=0$), equation (\ref{eigenvalue}) 
can be simplified as

\begin{eqnarray}
\mathrm{J_1}\left(\dfrac{\omega_n a}{v_T}\right) 
\left[2\dfrac{v_T}{v_L}\mathrm{J_1}\left(\dfrac{\omega_n a}{v_L}\right) -
\dfrac{\omega_n a}{v_T}\mathrm{J_0}
\left(\dfrac{\omega_n a}{v_L}\right)\right] = 0
\label{eigenvalue_inf}
\end{eqnarray}

\noindent It can be shown that the roots of $\mathrm{J_1}$ define 
modes with a purely axial displacement field. 
The roots of the second factor $g_n(a,v_L,v_T) = 
2v_T/v_L\mathrm{J_1}(\omega_n a / v_L) - \omega_n a / v_T\mathrm{J_0}
(\omega_n a / v_L)=0$ define the well-known breathing modes with a purely
radial displacement field. The displacement field distributions of some of
these modes are depicted in Fig.\ref{fig2}c. 

The frequency signature
obtained on a single free standing $200~\mathrm{nm}$ diameter copper nanowire 
with pump and probe superimposed is depicted in Fig.\ref{fig2}a. 
In an infinite wire, excitation and detection 
mechanisms essentially favor modes exhibiting large radial displacements \cite{Major2013}.
Taking 
advantage of the huge acoustic confinement, the signal exhibits a complex structure
composed of the fundamental breathing mode around $15.6~\mathrm{GHz}$, and its 
two first harmonics located at $39.6~\mathrm{GHz}$ and 
$60~\mathrm{GHz}$. Using this rich vibrational behavior, one can then 
expect to determine the three 
nanowire properties $\left(a,v_L,v_T\right)$ by solving this non linear 
inverse problem. However, it doesn't matter how many breathing mode frequencies 
are experimentally determined since the $g_n$ are homogeneous function of 
order zero : $g_n(\lambda a,\lambda v_L,\lambda v_T) = g_n(a,v_L,v_T)$ 
with $\lambda \neq 0$. When a solution is found, there is no unicity. 
One has to determine the dilatation factor $\lambda$ with an other 
set of measurements. For instance, $a$ can be determined using SEM or Atomic Force
Microscopy. Our SEM measurements give 
us a $200~\mathrm{nm}$ nanowire diameter which allows an estimation of the
velocity, equal to $v_L = 4.5\cdot 10^3~\mathrm{m \cdot s^{-1}}$ and 
$v_T = 2.2\cdot 
10^3~\mathrm{m \cdot s^{-1}}$ for longitudinal and transverse waves 
respectively, in good adequation with the copper elastic 
constants \cite{ledbetter1974elastic}. However, finding a single nano-object 
previously studied by femtosecond transient reflectometry under SEM often 
results in a very tedious and unfruitful task. Consequently, a means to 
characterize the material and geometrical properties of nanowires completely 
with picosecond acoustic measurement remains to be found.

In the following, we show that the determination of the three nanowire
properties $\left(a,v_L,v_T\right)$ can be 
adressed by the experimental
observation of propagating acoustic nanowaves corresponding to the 
$\mathrm{L(0,0)}$ and $\mathrm{L(0,2)}$ modes. It can be demonstrated that,
in the case of $\mathrm{L(0,0)}$, 
the first order development of the dispersion relation around $k=0$ 
is $\omega = v k$ with $v = \sqrt{E/\rho}$. We also 
demonstrated that the development of the dispersion equation
around $K=0$ of the first radial breathing mode $\mathrm{L(0,2)}$ that verifies 
$2v_L/v_T\mathrm{J_1}(\omega_2 a / v_L) - \omega_2 a / v_T\mathrm{J_0}
(\omega_2 a / v_L)=0$ is parabolic : $\omega a / v_L = 
\omega_2 a / v_L + \delta (ka)^2$ with

\begin{equation}
\delta  = \dfrac{4\left(2p_2\beta\mathrm{J_0}(p_2\beta)-
3\mathrm{J_1}(p_2\beta)\right)+p_2^2\beta^4\mathrm{J_1}(p_2\beta)}
{2p_2\mathrm{J_1}(p_2\beta)\left(p_2^2\beta^4+4(1-\beta^2)\right)}
\label{delta}
\end{equation}

\noindent where $p_2 = P_2(K=0) = \omega_2 a/v_L$. The time and space evolution
of a wave packet resulting from a propagation with linear and parabolic
dispersion relation is calculated, and the general
form $\omega(k) = \alpha k^2+\beta k+\gamma$ is chosen. Assuming a 
gaussian pump beam with diameter $\sigma = 0.5~\mathrm{\mu m}$ at $1/e^2$ 
and a uniform radial dilation of the nanowire at $z=0$ and $t=0$, the initial
deformation is $\eta(z) = 4/(\sigma\sqrt{2\pi}) 
\exp\left(-8z^2/\sigma^2\right)$. As we do not provide physical insight on the
excitation mechanism, we suppose no mode dependence in the excitation amplitude. 
According to this zero order approximation, each $k$ mode is excited with the same unitary 
amplitude, leading to $\hat{\eta}(k) = \int_{-\infty}^\infty \eta(z)
\exp\left(\mathrm{i}kz\right) \mathrm{d}z =
\exp\left(-k^2\sigma^2/32\right)$. We notice that only wave numbers verifying 
$k^2\sigma^2/32<2$, that is to say $k<20~\mathrm{\mu m^{-1}}$, are excited with 
significant amplitude. 
Such a low $k$ amplitude should lead to a small $k$ dependance in the 
excitation amplitude of each mode 
which justifies the above approximation. Each $k$ mode propagates as a 
plane-wave $\exp\left(\mathrm{i}(kz-\omega(k)t)\right)$ along the $z$ axis 
of the wire resulting in the following gaussian wave packet 

\begin{eqnarray}
\psi(z,t) & = & \int_{-\infty}^\infty \hat{\eta}(k)
\exp\left(\mathrm{i}(kz-\omega(k)t)\right) \mathrm{d}k\\
	& = & \dfrac{\sqrt{32\pi}}{\sqrt{\sigma^2+32i\alpha t}}
\exp\left(-\dfrac{8(\beta t-z)^2}{\sigma^2+32\mathrm{i}\alpha t}\right)
\exp(-\mathrm{i}\gamma t)
\label{paquet}
\end{eqnarray}

\noindent As the detection is achieved at $z=z_0$ using a gaussian probe beam, 
with diameter $\nu = 1~\mathrm{\mu m}$ at $1/e^2$, a convolution with 
$\phi(z) = 4/(\nu\sqrt{2\pi}) \exp\left(-8(z-z_0)^2/\nu^2\right)$ is applied. 
Finally, the experimental signal  
at a distance $z_0$ from the pump beam may be assumed to be proportional to the
real part of the following equation

\begin{eqnarray}
\Delta r(t) & = & \int_{-\infty}^\infty \phi(z)\psi(z,t) \mathrm{d}z \\
			& = & \dfrac{\sqrt{32 \pi}\exp(-\mathrm{i}\gamma t)}
{\sqrt{\nu^2+\sigma^2+32\mathrm{i}\alpha t}} \exp
{\left(-\dfrac{8(z_0-\beta t)^2}{\nu^2+\sigma^2+32\mathrm{i} \alpha t}\right)}
\label{reflec}
\end{eqnarray}

\noindent We can also extract the envelop of the gaussian wave packet with 
the modulus of $\Delta r(t)$. In the following, we will discuss the experimental
 results in the light of this analytical expression. Considering the 
$\mathrm{L(0,0)}$ mode, we will take $\alpha = 0$, $\beta = \sqrt{E/\rho}$, 
$\gamma = \omega_0 = 0$. Considering the $\mathrm{L(0,2)}$ mode, 
the parameter values will be $\alpha = \delta v_L a$, $\beta = 0$, $\gamma = 
\omega_2$.

\begin{figure}
  \includegraphics{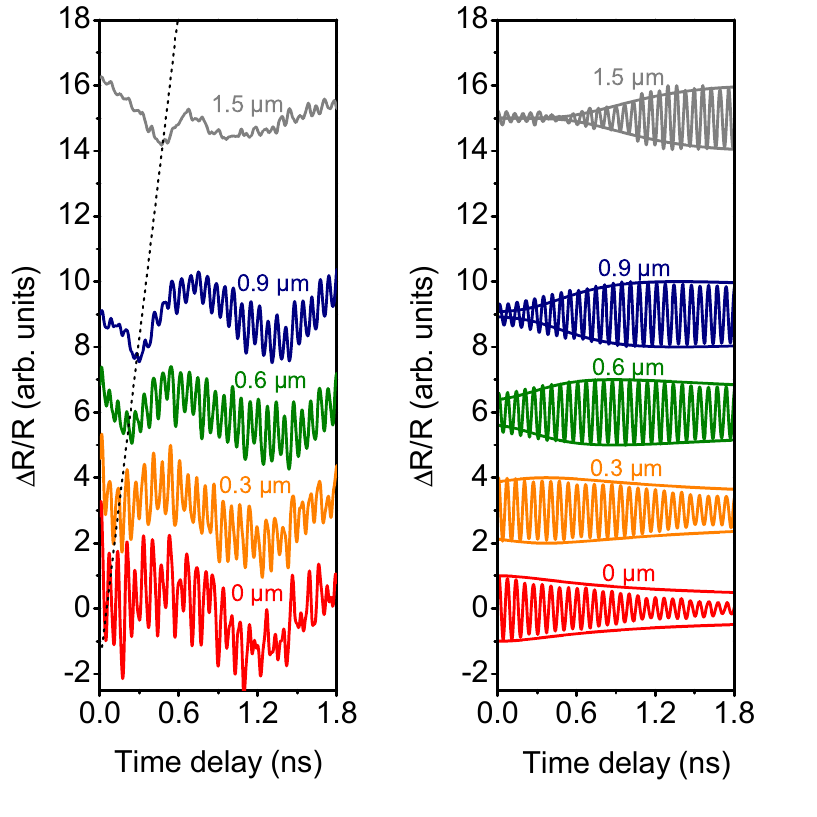}
  \caption{(Color online) Left (a) : Raw data of the transient reflectivity 
obtained on a single free standing $200~\mathrm{nm}$ diameter copper nanowire. 
The distance labeled above each curve is the pump-probe separation $z_0$. 
The high-frequency signal is the $15.6~\mathrm{GHz}$ first radial breathing 
mode. The additional feature that appears at $600~\mathrm{nm}$ with a probe 
delay of $170~\mathrm{ps}$ and at $1500~\mathrm{nm}$ with a probe delay of 
$400~\mathrm{ps}$ is the signature of the $\mathrm{L(0,0)}$ mode : the black
dotted line follows the maximum of the deformation and demonstrates a linear propagation. Right (b) : 
Same signal normalized to $[-1,1]$ with a second order band-pass filter centered 
at the first radial breathing mode frequency of $15.6~\mathrm{GHz}$. We thus highlight the dispersion of 
the first breathing mode. The solid line envelop is $\left|\Delta r(t)\right|$ 
calculated analytically with $\alpha = 1.7 \cdot 10^{-4}~\mathrm{m^2 
\cdot s^{-1}}$, $\beta = 0$, $\gamma = 15.6~\mathrm{GHz}$, $\nu = 
1~\mathrm{\mu m}$ and $\sigma = 0.5~\mathrm{\mu m}$. A comparison between 
non-filtered Fig.~\ref{fig3}a and filtered Fig.~\ref{fig3}b shows 
that there is no artifact added by such filtering process.”}
  \label{fig3}
\end{figure}

\begin{figure}
  \includegraphics{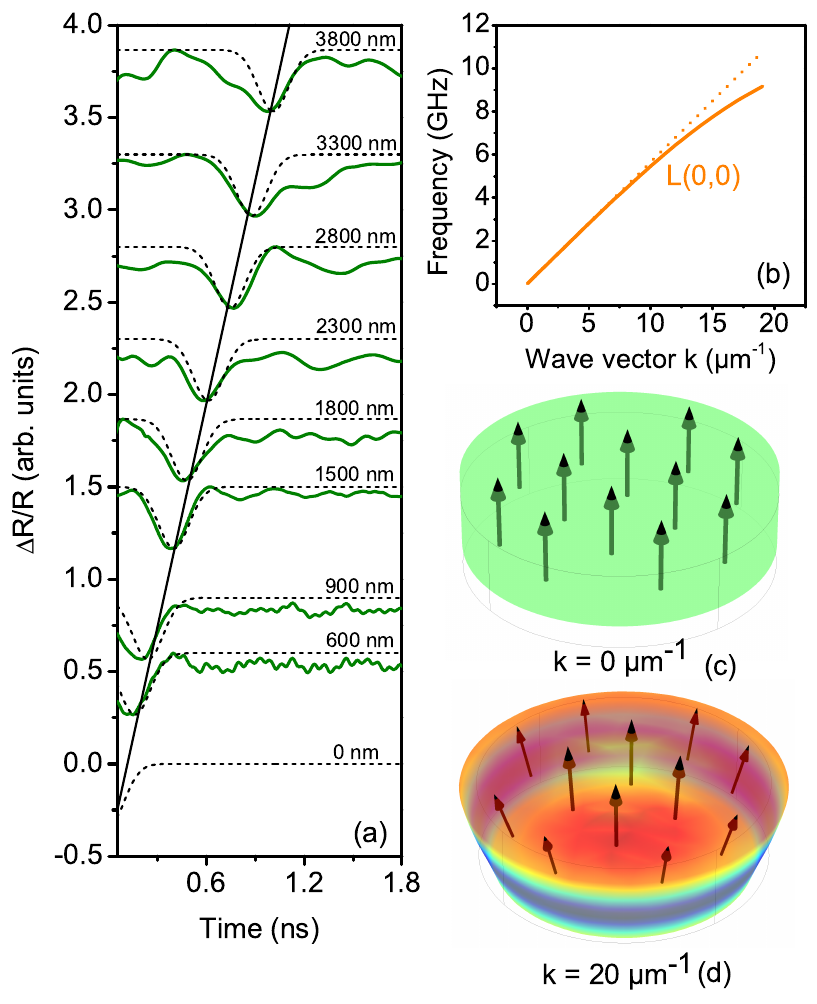}
  \caption{(Color online) (a) : Transient reflectivity signal obtained on a 
single free standing $200~\mathrm{nm}$ diameter copper nanowire with a 
$200~\mathrm{ms}$ time integration constant that acts as a hardware low-pass filter. 
Signals are normalized to $[-0.3,0]$. The distance label above 
each curve corresponds to the pump-probe separation $z_0$. The dotted line 
is $\left|\Delta r(t)\right|$ calculated analytically with $\alpha = 0$, 
$\beta = 3.6\cdot 10^{3}~\mathrm{m \cdot s^{-1}}$, $\gamma = 0$, 
$\nu = 1~\mathrm{\mu m}$ and $\sigma = 0.5~\mathrm{\mu m}$. The black solid 
line follows the maximum of the deformation and demonstrates a linear 
propagation (b) : Dispersion 
relation of the $\mathrm{L(0,0)}$ branch. The solid line corresponds to the
exact relation, the dotted line is the linear approximation $\omega = 
\sqrt{E/\rho} k $. (c) : The displacement field distribution of mode 
$\mathrm{L(0,0)}$ at infinite wavelength calculated with three dimensional 
FEM exhibits no radial displacement. (d) : The displacement 
field distribution of mode $\mathrm{L(0,0)}$ at 
$k \approx 20~\mathrm{\mu m^{-1}}$ calculated with three dimensional 
FEM shows a significant radial displacement.}
  \label{fig4}
\end{figure}

In order to be sensitive to the propagation phenomenon, the pump and probe beams
have to be separated using an experimental setup which allows to tilt the
probe before the last objective lens \cite{Kelf}. The transient reflectivity 
obtained on a single free standing $200~\mathrm{nm}$ diameter copper nanowire, with 
pump-probe separation ranging from $z_0 = 0~\mathrm{nm}$ to 
$z_0 = 1.5~\mathrm{\mu m}$, is presented in Fig.~\ref{fig3}. The distance 
labeled above each curve is the pump-probe separation $z_0$. At small 
pump-probe separation, the signal is mainly composed of a high-frequency
signature at
$15.6~\mathrm{GHz}$ which corresponds to the first radial breathing mode. As the 
pump-probe separation increases, this high-frequency wave-packet shifts to a
longer time delay as expected for a propagation guided along the nanowire axis. 
However, this signal suffers strong attenuation which 
results in a poor signal-to-noise ratio at large pump-probe separations. 
To get rid of the attenuation and to increase the signal-to-noise ratio at 
large pump-probe separation, the signal is normalized to $[-1,1]$ and a 
band-pass filter centered at $15.6~\mathrm{GHz}$ is applied. The resulting 
signal is ploted in Fig.~\ref{fig3}b. The strong dispersion of this mode 
is clearly revealed. The solid line envelop $\left|\Delta r(t)\right|$ 
calculated analytically with a purely parabolic dispersion relation  : 
$\alpha = 1.7 \cdot 10^{-4}~\mathrm{m^2 \cdot s^{-1}}$, $\beta = 0$, 
$\gamma = 15.6~\mathrm{GHz}$, $\nu = 1~\mathrm{\mu m}$ and 
$\sigma = 0.5~\mathrm{\mu m}$, fits very well with the whole set of experimental
results. It is remarkable that the extracted value
$\alpha$ is the numerical result of $\delta v_L a$ with $v_L = 4.5 
\cdot 10^3~\mathrm{m \cdot s^{-1}}$, $v_T = 2.2 \cdot 10^3~\mathrm{m 
\cdot s^{-1}}$ and $a=100~\mathrm{n m}$, which are the values obtained 
experimentally using the 
breathing mode at $K=0$ and the SEM measurements. Furthermore, a close inspection of figure \ref{fig3}
reveals that a signal deformation appears at $z_0 = 600~\mathrm{nm}$ 
with a probe 
delay of $170~\mathrm{ps}$ and propagates through $z_0 = 1500~\mathrm{nm}$ 
with a probe delay of $400~\mathrm{ps}$. This signature can be associated to the 
$\mathrm{L(0,0)}$ mode. It is interesting to note that the displacement field 
distribution of mode $\mathrm{L(0,0)}$ at an infinite wavelength 
calculated with three dimensional FEM presents no radial displacement explaining
 its absence when the pump and probe are superimposed.
On the contrary, the displacement field distribution of mode $\mathrm{L(0,0)}$ 
at $k \approx 20~\mathrm{\mu m^{-1}}$ calculated with three dimensional 
FEM shows a significant radial displacement. Therefore, this mode will be more
easily detected with small pump and probe beams which provide a large wave
vector distribution. We are then able to detect 
this $\mathrm{L(0,0)}$ mode when the pump-probe spatial separation inscreases.
To further investigate this signal, we explore
larger pump-probe spatial separation as reported in Fig.~\ref{fig4}.

This transient reflectivity signal is obtained on a 
single free standing $200~\mathrm{nm}$ diameter copper nanowire with a 
$200~\mathrm{ms}$ time integration constant that acts as a low-pass filter and 
increases the signal-to-noise ratio dramatically, which is much needed at
distances as large as $3.8~\mathrm{\mu m}$. 
Signals are normalized to $[-0.3,0]$ to get rid of the attenuation which 
is not taken into acount in our theorical model. The distance label above 
each curve corresponds to the pump-probe separation $z_0$. The dotted line 
is $\left|\Delta r(t)\right|$ calculated analytically with $\alpha = 0$, 
$\beta = 3.6\cdot 10^{3}~\mathrm{m \cdot s^{-1}}$, $\gamma = 0$, 
$\nu = 1~\mathrm{\mu m}$ and $\sigma = 0.5~\mathrm{\mu m}$. Such a
best fit value is consistant with $\sqrt{E/\rho}$. Unlike the complex expression of
the parabolic coefficient $\delta$ of the $\mathrm{L(0,2)}$ mode and the
resulting complex spreading of the gaussian wave-packet in Fig.~\ref{fig3}, this 
$\mathrm{L(0,0)}$ mode provides a unique way to get the Young modulus of a
material independently of any geometrical dimensions. Simple and precise reading
 of the deformation maximum's propagation velocity gives direct access to 
the Young modulus. Finally, measuring $v = 3.6\cdot 10^{3}~\mathrm{m 
\cdot s^{-1}}$, $f_2 = 15.6~\mathrm{GHz}$ and $f_5 = 39.6~\mathrm{GHz}$ 
gives : $a=91 \pm 3~\mathrm{nm}$, $v_L=4.1 \pm 0.2 \cdot 10^3~\mathrm{m 
\cdot s^{-1}}$ and $v_T =2.2 \pm 0.1 \cdot 10^3~\mathrm{m \cdot s^{-1}}$. These
values are in good agreement both with copper elastic constants and SEM
measurements.

In summary, we have investigated the propagation of two guided modes
 in single free standing copper nanowires. By considerably reducing the
 relaxation channel towards the substrate, the suspended nanowires provide a
 unique tool to observe the propagation of gigahertz coherent acoustic waves with
spatial separations between generation and detection as large as 
$4~\mathrm{\mu m}$. A rigorous approach allows us to assign the main components of 
the reflectometry signal to a precise acoustic mode. These experimental 
observations also lead to two equations that accurately complete the set of 
equations at infinite wavelength thus allowing the determination of 
elastic properties as well as the geometrical radius $a$ of the nanowire. 
Beyond this new way to achieve material and geometrical characterization 
through picosecond acoustic experiment, this study
paves the way to use nanowires as nanometric lateral size acoustic transducers.


\bibliography{bibliography}

\end{document}